\documentclass[twocolumn]{article}

\usepackage{fullpage}
\usepackage{latexsym}
\usepackage{amsfonts}
\usepackage{amsmath}
\usepackage{hyperref}
\newcommand{\ket}[1]{\left| #1 \right\rangle }
\newcommand{\bra}[1]{\left\langle #1 \right| }

\newtheorem{theorem}{Theorem} 
\newtheorem{lemma}[theorem]{Lemma}

\newtheorem{definition}{Definition} 

\newenvironment{proof}{\begin{description} \item[Proof:] }{$\Box$ \end{description}}
\newenvironment{remark}{\begin{description} \item[Remark:] }{\end{description}}

\begin{document}
\title{Quantum secret sharing for general access structures}
\date{May 1998; revised January 1999} \author{Adam
  Smith\\MIT\thanks{Work done while author was at McGill University,
    Montreal. Supported by an NSERC undergraduate research
    grant.}\\{\tt asmith@theory.lcs.mit.edu}}

 \maketitle

\begin{abstract}
  We explore the conversion of classical secret-sharing schemes to
  quantum ones, and how this can be used to give efficient
  \textsc{qss} schemes for general adversary structures.  Our first
  result is that quantum secret-sharing is possible for any structure
  for which no two disjoint sets can reconstruct the secret (this was
  also proved, somewhat differently, in \cite{Gott99}). To obtain this
  we show that a large class of \emph{linear} classical \textsc{ss}
  schemes can be converted into quantum schemes of the same
  efficiency.
  
  We also give a necessary and sufficient condiion for the direct
  conversion of classical schemes into quantum ones, and show that all
  group homomorphic schemes satisfy it.
\end{abstract}

\section{Introduction}
\label{sec:intro}

A classical secret sharing scheme is a (usually) randomized encoding
of a secret $s$ into a $n$-tuple, the coordinates of which are each
given to different players in the player set $P$. The encoding is a
secret sharing scheme if there exists a collection ${\cal A}$ of
subsets of $P$ (called the \emph{adversary structure}) such that no
set of players in ${\cal A}$ gets any information about $s$ from their
shares, but any set of players not in ${\cal A}$ will be able to
compute $s$. The classic example of this is due to Shamir
\cite{Shamir79}. He gives a construction based on polynomials over a
finite field of a \emph{threshold} secret-sharing scheme for any
threshold $t$ and any number of players (in such a scheme, ${\cal
  A}=\{ B \subseteq P : |B|\leq t\}$).

%Note that for the scheme to be perfectly hiding, the secret must always come
%from a pre-determined finite secret space ${\cal S}$ (this is easy to
%see: if ${\cal S}$ were infinite, the mere size of the shares given to
%each participant would give away information on the secret).

The idea of sharing \emph{quantum} secrets was first described and
solved for the case $t=1,n=2$ by Hillery \emph{et al.}~in
\cite{HBB99}\footnote{In fact, \cite{HBB99} shows how efficency can be
  gained in the insecure channels model by combining the key
  distribution and secret-sharing layers of the protocol. An even more
  efficient protocol was suggested in \cite{KKI99}.}. A more general
solution, for all $t > \frac{n}{2}-1$, was recently given by Cleve
\textsl{et al.}~(CGL, \cite{CGL99}). Their scheme is a direct
generalization of the well-known Shamir scheme \cite{Shamir79}, with
all calculations done unitarily and ``at the quantum level'',
\textsl{i.~e.}~replacing random choices with equal superpositions over
those choices.

In next section we give definitions and background.  In section 3, we
then prove that classical \emph{linear} secret-sharing schemes, with
an appropriate adversary structure, can be converted into quantum
schemes with the same complexity, both in terms of share size and
encoding/reconstruction. This gives another proof of theorem
8 from \cite{Gott99}. In the last section, we give a necessary and
sufficient condition for (not necessarily linear) classical
\textsc{ss} schemes to become quantum ones when run at the quantum
level, and observe that all group homomorphic schemes obey this
condition.

\section{Preliminaries}
\label{sec:genstruct}

\subsection{Adversary structures}
\label{sec:struct}

Given a set of players $P$, an adversary structure ${\cal A}$ over $P$
is a set of subsets of players which is downward-closed under
inclusion:

$$(B \in {\cal A} \textrm{ and } B' \subseteq B) \Longrightarrow B' \in {\cal A}.$$

Normally such a structure is used to represent the collection of all
coalitions of players which a given protocol can tolerate without
losing security: as long as the set of cheating players is in ${\cal
  A}$, the cheaters cannot breach the security of the protocol.

Secret-sharing schemes usually tolerate \emph{threshold structures},
which are of the form ${\cal A} = \{ B \subseteq P : |B| \leq t \}$ for some
$t$. However, when working with more general structures, the following
definitions prove useful.

\begin{definition}
  An adversary structure ${\cal A} \subseteq 2^P$ is ${\cal Q}^2$ if
  no two sets in ${\cal A}$ cover $P$, that is
  $$\not \exists B_1,B_2 \in {\cal A} : \quad B_1 \cup B_2 = P.$$
\end{definition}

\begin{definition}
  The \emph{dual} of an adversary structure ${\cal A}$ over $P$ is the
  collection

  $${\cal A}^* = \{ B \subseteq P: B^c \notin {\cal A}\}$$
  
  where $B^c$ denotes the complement $P-B$.
\end{definition}

\begin{definition}
  A structure ${\cal A}$ over $P$ is ${\cal Q}^{2*}$ if its dual
  ${\cal A}^*$ is ${\cal Q}^2$. This means that any two sets not in
  ${\cal A}$ will have a non-empty intersection.
\end{definition}

It is interesting to note that ${\cal A}$ is ${\cal Q}^2$ iff ${\cal
  A} \subseteq {\cal A}^*$. Dually, ${\cal A}$ is ${\cal Q}^{2*}$ iff
${\cal A} \supseteq {\cal A}^*$. Consequently, a collection is
\emph{self-dual} iff it is both ${\cal Q}^2$ and ${\cal Q}^{2*}$.

\subsubsection{Monotone functions}
\label{sec:mf}

%\begin{definition}
%  For partial orders $\leq$ on sets $A$ and $B$, we say that a function
%  $f: A \to B$ is \emph{monotone} if for $x,y \in A$ we have
%  $$x \leq y \Longrightarrow f(x) \leq f(y)$$
%\end{definition}

We can define a partial order on $\{ 0,1 \} ^n$ by the rule ``${\bf x}
\leq {\bf y}$ iff each coordinate of ${\bf x}$ is smaller than the
corresponding coordinate of ${\bf y}$.'' 

By identifying $\{ 0, 1\} ^n$ with $2^{\{ 1,\ldots ,n \}}$, the
relation $\leq$ on $\{ 0,1 \} ^n$ corresponds to inclusion
($\subseteq$) in $2^{\{ 1,\ldots ,n \} }$. Then a monotone function
$f$ corresponds to a function from $2^{ \{ 1,\ldots ,n \} }$ to $\{
0,1 \}$ such that $A \subseteq B \Longrightarrow f(A) \leq f(B)$.

Such a monotone function $f$ naturally defines an adversary structure
${\cal A}_f = f^{-1}(\{0\}) = \{ B \subseteq P: f(B)=0 \}$. Moreover,
$f$ is called ${\cal Q}^2$ (resp. ${\cal Q}^{2*}$) iff ${\cal A}_f$ is
${\cal Q}^2$ (or ${\cal Q}^{2*}$).

\subsection{Monotone span programs}
\label{sec:msp}

Span programs were introduced as a model of computation in
\cite{KW93}. They were first used for multiparty protocols in
\cite{CDM98} under this name, although a similar construction,
attributed to Brickell, already existed (\cite{StinsonBook11}). In
this section we define some concepts related to monotone span
programs.

\begin{definition}
  A {\bf {monotone span program}} (MSP) over a set $P$ is a triple
  $(K,M,\psi)$ where $K$ is a finite field, $M$ is a $d \times e$
  matrix over $K$ and $\psi: \{1, \ldots, d \} \to P$ is a function
  which effectively labels each row of $M$ by a member of $P$.
\end{definition}

The MSP associates to each subset $B \subseteq P$ a subset of the rows of
$M$: the set of rows $l$ such that $\psi(l) \in B$. This corresponds to a
linear subspace of $K^e$ (the span of those rows).  The monotone
function $f: 2^P \to \{ 0,1\}$ defined by a MSP is given by the rule
``$f(B)=1$ if and only if the target vector $\epsilon = (1,0,0,\ldots,0)$ is in
the subspace associated with $B$''.  If we denote by $M_B$ the
submatrix of $M$ formed of the rows $l$ such that $\psi (l) \in B$ then
we get that

\begin{displaymath}
  f(B) = 1 \iff \epsilon \in Im(M_B^T).
\end{displaymath}

In fact, given any monotone function $f$, we can construct a MSP which
computes it. The size of the MSP will be at most proportional to the
size of the smallest monotone threshold formula for $f$, but may in
some cases be exponentially smaller \cite{BabaiGKRSW96,CDM98}.

The proof uses the following fact from linear algebra.  Here the dual
of a vector subspace $W$ is denoted $W^\perp=\{{\bf u}: {\bf u}^\top {\bf
  w} = 0 \quad \forall {\bf w} \in W\}$.

\begin{remark}
  Denote the dual of a vector subspace $W$ by $W^\perp=\{{\bf u}: {\bf
    u}^\top {\bf w} = 0 \quad \forall {\bf w} \in W\}$. For any matrix $M$ we
  have $Im(M^\top) = ker(M)^\perp$. Thus, $f(B)=0$ iff $\exists {\bf v}: M_B
  {\bf v} = {\bf 0}$ and $\epsilon^\top{\bf v} \neq 0$.
\end{remark}

\subsubsection{Secret-sharing from MSP's}
\label{sec:ss-msp}

Given a MSP $(K,M,\psi)$, we can define a classical secret sharing
scheme which tolerates the adversary structure ${\cal A}_f$ induced by
the MSP. Say the dealer has a secret $s \in K$. He extends it to an
$e$-rowed vector by adding random field elements $a_2,\ldots,a_e$ to make
a vector ${\bf s_*} = (s,a_2,\ldots,a_e)$. The dealer gives the $l$th
component of ${\bf {\hat s}} = M {\bf s_*}$ to player
$P_{\psi(l)}$.  If ${\bf {\hat s}}_A$ denotes the elements of
${\bf {\hat s}} $ with indices in $A$ where $A \subseteq \{ 1,\ldots , d\}$, then
each $P_i$ receives ${\bf {\hat s}}_{\psi^{-1}(i)}$. 

The \textsc{ss} scheme thus defined tolerates exactly the adversary
structure ${\cal A}_f$. 

Note that the concept of MSP's is very general: any linear
secret-sharing scheme (i.e. one in which the encoding of the secret is
given by a linear map over a field) can be formulated as a MSP-based
scheme \cite{CDM98}.  The Shamir scheme is a special case, where $M$
is a $n \times (k+1)$ Vandermonde matrix, $e=k+1$, $d=n$, and $\psi$ is the
identity on $\{1,\ldots,n\}$.

\subsection{Secret sharing with general access structures}
\label{sec:qss}

With classical data, secret sharing is possible for any access
structure. Given a monotone threshold formula for a function $f$,
Benaloh and Leichter \cite{BL88} gave a construction for ${\cal A}_f$
with efficency proportional to the size of the formula. This is
improved on by constructions based on monotone span programs (section
\ref{sec:ss-msp}), which are always at least as efficient as the
Benaloh-Leichter scheme but can be super-polynomially more so.

When sharing quantum data, the situation is slightly different.
Because of the no-cloning theorem, it is impossible to share secrets
with an adversary structure which is not ${\cal Q}^{2*}$ (since then
one can find two disjoint sets which can reconstruct the secret based
on their shares). Because a pure-state \textsc{qss} scheme is also a
quantum code correcting erasures on the sets described by its
adversary structure, we also get that any pure-state \textsc{qss}
scheme has an adversary structure which is in fact self-dual
\cite{CGL99}. The natural converse to this is

\begin{theorem}
  \label{thm:main-qss}
  Given any ${\cal Q}^{2*}$ structure ${\cal A}$, we can find a
  \textsc{qss} scheme for ${\cal A}$. If ${\cal A}$ is self-dual, then
  the scheme can be a pure-state one.
\end{theorem}

This was proved for the case of threshold structures in \cite{CGL99}:
their construction works when the number of cheaters $t$ is more than
$\frac{n}{2}-1$ (\textsl{i.~e.}~it takes more than $\frac{n}{2}$
players to reconstruct the secret). Moreover, theirs is a pure-state
scheme when $n=2t+1$ (these correspond to the ${\cal Q}^{2*}$ and
self-dual conditions, respectively).

The full theorem was stated but not proved in \cite{CGL99}. We give a
proof here, based on monotone span programs. Another proof, due to
Daniel Gottesman and based on purification of quantum superoperators,
appeared in \cite{Gott99}.

\section{Quantum secret-sharing from classical linear schemes}
\label{sec:ss-app}

We assume that the reader is familiar with the notation and basic
concepts of quantum computing. For clarity, we will ignore
normalization factors.

\subsection{Pure-state linear QSS}
\label{sec:lin-qss}

Cramer \textsl{et al.}~\cite{CDM98} pointed out that any linear
secret-sharing scheme can be realized as a MSP-based scheme. In this
section, I show that any MSP with adversary structure ${\cal A}$ gives
rise to a quantum erasure-correcting code for erasures occuring on any
set of positions in ${\cal A} \cap {\cal A}^*$. In the case where ${\cal
  A}$ is self-dual, this yields a pure-state quantum secret-sharing
scheme for ${\cal A}$.

The idea is the same as that for the CGL scheme \cite{CGL99}. First
choose a MSP, say $(K,M,\psi)$. Note that \textsc{wlog} all $e$ rows
of $M$ are linearly independent and so we can extend $M$ to an
invertible $d \times d$ matrix $M'$. We can construct a quantum
circuit $\tilde{M}$ implementing multiplication by $M'$ and thus
encode a basis state $\ket{s}$, for $s \in K$, as

\begin{eqnarray*}
& & \tilde{M} \left( \ket{s} \otimes \sum_{{\bf a} \in K^{e-1}} \ket{a_1 \cdots
    a_{e-1}} \otimes \ket{0\cdots 0} \right)  \\
& & = \sum_{{\bf a} \in K^{e-1}}
\ket{M {s \choose {\bf a}}}
\end{eqnarray*}

{\small (The expression ${s \choose {\bf a}}$ denotes the column vector
obtained by adjoining $s$ to the beginning of the vector {\bf a}).}

This scheme can be extended by linearity to arbitrary states $\ket{\phi}
= \sum_{s \in K} \alpha_s \ket{s}$. The pieces of the encoded state are then
distributed according to the function $\psi$. We have:

\begin{theorem}
  \label{thm:qss-util}
  Let $(K,M,\psi)$ be a MSP with a.s. ${\cal A}$. Then the encoding
  above is corrects erasures on any set of positions in ${\cal A} \cap
  {\cal A}^*$.
\end{theorem}

To prove this, we need to show for any set $B$ which is in ${\cal A}$
but whose complement is not, the players in $A$ can reconstruct the
encoded data. We give a reconstruction procedure. The proof consists
of the two following lemmas.

First we show the existence of certain vectors used in the
reconstruction process.

\begin{lemma} \label{lem:u}
  Let $(K,M,\psi)$ be a MSP with a.s. ${\cal A}$. Suppose $B \in {\cal A}
  \cap {\cal A}^*$ (\textsl{i.e.}~$A=P-B$ is in ${\cal A}$).  Then there
  exists an invertible linear transformation $U$ on the shares of $A$
  such that after the transformation,
  \begin{enumerate}
  \item the first share contains the secret $s$;
  \item all remaining shares, including those of players in $B$, are
    distributed independently of $s$ when the $e-1$ other components
    of ${\bf s_*}$ are chosen at random.
  \end{enumerate}
\end{lemma}

\begin{proof}
  Say $A$ contains $m$ shares. Then we must construct $m$ linearly
  independent vectors ${\bf u}_1,{\bf u}_2,\ldots,{\bf u}_m$ such that
  \begin{enumerate}
  \item ${\bf u}_1^\top M_A {s \choose {\bf a}}= s$;
  \item If $U'$ is the matrix with rows given by ${\bf u}_2,\ldots,{\bf
      u}_m$, then the value $${U'M_A \choose M_B } {s \choose {\bf
        a}}$$
    is distributed independently of $s$.
  \end{enumerate}
  
  To satisfy the first condition, pick any ${\bf u}_1$ such that ${\bf
    u}_1^\top M_A = \epsilon^\top$. Such a vector must exist since by hypothesis
  the players in $A$ can reconstruct the secret.
  
  To satisfy the second condition, it's enough to ensure there exists
  ${\bf v}$ such that ${U'M_A \choose M_B } {\bf v} = {\bf 0}$ and
  $\epsilon^\top{\bf v} \neq 0$ (see section \ref{sec:msp}).
  
  Since $B \in {\cal A}$, we know that there is a ${\bf v}$ such that
  $\epsilon^\top{\bf v} \neq 0$ and $M_B {\bf v} = {\bf 0}$. Furthermore, the
  subspace $W=\{ {\bf u} \in K^m: {\bf u}^\top M_A {\bf v}=0\}$ has
  dimension $m-1$, and ${\bf u}_1$ is not in that space since ${\bf
    u}_1^\top M_A {\bf v} = \epsilon^\top {\bf v} \neq 0$. Hence any basis $\{{\bf
    u}_2,\ldots,{\bf u}_m\}$ of $W$ will do.
  
  The matrix $U$ whose rows are given by the $\textbf{u}_i$'s gives
  the desired transformation. Note that the $U$ doesn't depend on
  $\textbf{a}$.
\end{proof}

Finally we show that the reconstruction process works: 

\begin{lemma}
  Let $(K,M,\psi)$ be a MSP and let $B \in {\cal A} \cap {\cal A}^*$,
  $A=P-B$. Suppose a quantum state $\ket{\phi} = \sum_{s \in K} \alpha_s
  \ket{s}$ is encoded as described at the beginning of this section.
  Then the shares in $A$ can be used to reconstruct $\ket{\phi}$.
  Consequently, no information on $\ket{\phi}$ can be obtained from the
  shares in $B$.
\end{lemma}

\begin{proof}
  Consider the case when $\ket{\phi}=\ket{s}$ for some $s\in K$. Then the 
  encoded state can be written 
  
  $$\sum_{{\bf a}\in K^{e-1}} \ket{M_A {s \choose {\bf a}}} \ket{M_B {s
      \choose {\bf a}}}$$
  
  Construct a quantum circuit for the map ${\bf b} \longmapsto U {\bf b}$,
  where $U$ is constructed as in lemma \ref{lem:u}.  Denote by $U'$
  the matrix obtained by removing the first row of $U$.  Applying the
  circuit for $U$ only to the components of the encoded state
  corresponding to $A$, we get
  
  \begin{multline*}
 \sum_{{\bf a} \in K^{e-1}} \ket{U M_A {s \choose {\bf a}}} \ket{M_B
    {s \choose {\bf a}}} \\
 = \ket{s} \otimes \sum_{{\bf a}\in K^{e-1}} \ket{U'
    M_A {s \choose {\bf a}}} \ket{M_B {s \choose {\bf a}}}
\end{multline*}
  
  However, by construction the joint distribution of $U' M_A {s
    \choose {\bf a}}$ and $M_B {s \choose {\bf a}}$ is independent of
  $s$ when $\textbf{a}$ is chosen uniformly at random (lemma
  \ref{lem:u}). Hence, for an arbitrary state $\ket{\phi}$ this
  procedure yields
  
  $$\ket{\phi} \otimes \sum_{{\bf a}\in K^{e-1}} \ket{U' M_A {0 \choose {\bf
        a}}} \ket{M_B {0 \choose {\bf a}}}$$
  
  By a strong form of the no cloning theorem, the correctness of the
  reconstruction implies that the shares of $B$ give no information at
  all on $\ket{\phi}$.
\end{proof}

(This completes the proof of theorem \ref{thm:qss-util}).

When the adversary structure ${\cal A}$ defined by a MSP is ${\cal
  Q}^2$, we have ${\cal A} \subseteq {\cal A}^*$.  Hence, the previous
theorem shows that erasures on any set of coordinates in ${\cal A}$
can be corrected. In addition, if ${\cal A}$ is self-dual
(\textsl{i.~e.}~both ${\cal Q}^2$ and ${\cal Q}^{2*}$) then the
qualified sets are precisely the complements of sets in ${\cal A}$ and
hence every qualified set can reconstruct the secret but no
unqualified set gets any information on it. Thus we have shown theorem
\ref{thm:main-qss} for the case of self-dual structures.

\subsection{Mixed-state linear QSS}
\label{sec:qss-msp-mix}

To handle structures which are simply ${\cal Q}^{2*}$, we follow the
strategy of \cite{CGL99}: first extend to a self-dual structure and
then ``trace-out'' the new share(s).

To extend a structure ${\cal A}$ over a player set $P$, add a new
player to $P$ (say $\tau$):

\begin{lemma}
  For any ${\cal Q}^{2*}$ adversary structure ${\cal A}$ over a player
  set $P$, the structure ${\cal A}'$ over the set $P'=P \cup \{ \tau \}$
  given by
  $${\cal A}' ={\cal A} \cup \big\{ B \cup \{ \tau \} : B \in {\cal A}^*
  \big\}$$
  is self-dual and its restriction to $P$ yields ${\cal A}$.
\end{lemma}

\begin{proof} Elementary, using the fact that \\
  ${\cal A}$ is ${\cal Q}^{2*} \iff {\cal A}^* \subseteq {\cal A}$.
\end{proof}

Thus, a pure-state QSS scheme for ${\cal A}'$ will yield a mixed-state
scheme for ${\cal A}$ by throwing out the share corresponding to $\tau$.
For the construction to be efficient, we need the following:

\begin{lemma}
  Given a MSP for ${\cal A}$, an MSP for ${\cal A}'$ can be
  efficiently constructed.
\end{lemma}

\begin{proof}
  Note that the new access structure is $\Gamma ' = \Gamma \cup \{ B \cup \{\tau\} :
  B \in \Gamma^*\}$ (here $\Gamma, \Gamma^*, \Gamma'$ are the complements of ${\cal A},
  {\cal A}^*, {\cal A}'$ resp.). Thus if $f,f^*,f'$ are functions
  detecting membership in ${\cal A},{\cal A}^*,{\cal A}'$
  respectectively, and if $f_\tau$ detects the presence of $\tau$ in a
  set, then $f' = f \lor (f^* \land f_\tau)$.
  
  Now to construct the desired MSP, first obtain an MSP for ${\cal
    A}^*$ according to \cite{Fehr99}.  The MSP for $A'$ can then be
  constructed by composition from MSP's calculating \textsc{and} and
  \textsc{or}.
\end{proof}

The resulting MSP is at most a constant times the size of the
original.

\section{QSS from classical SS}
\label{sec:qss-css}

%In \cite{CC97}, Cerf and Cleve show some essential differences between
%information-theoretic characterizations of classical and quantum
%erasure-correcting codes. Implicit in their discussion is the fact
%that quantum erasure-correcting codes are more closely related to
%classical secret-sharing schemes than to classical erasure-correcting
%codes, provided we take the qualified sets to be the complements of
%non-qualified sets.

A natural conjecture given the results of the previous section is that
\emph{any} classical secret-sharing scheme for an adversary structure
will give a quantum erasure-coorecting for erasures in ${\cal A} \cap {\cal A}^*$. I show here a condition on the scheme for this to be the
case. Not all schemes satisfy the condition, though a large class of
them does, in particular group-homomorphic ones.

The corollary to this, as before, is that when ${\cal A}$ is
self-dual, the resulting quantum scheme is a \textsc{qss} scheme for
${\cal A}$. Note that the main difference between the proof we give
here and that of the preivous section is that here we don't guarantee
that the reconstructon procedure is efficient, only that it exists and
is unitary.

\subsection{A general condition}
\label{sec:cond}

A classical secret sharing scheme can be thought of as a probabilistic
map $E$ from a secret space ${\cal S}$ into $n$ ``share spaces''
${\cal Y}_1,\ldots,{\cal Y}_n$. The random input can be modeled as a
choice from some set ${\cal R}$ with a given probability distribution.
Now consider some set $U \in {\cal A} \cap {\cal A}^*$ and let $Q=U^c$ be
its complement ($Q$ is qualified). Let $S$ be the random variable
corresponding to the secret and let $Y_u$ and $Y_q$ be those
corresponding to the shares in $U$ and $Q$ respectively. Denote their
concatenation $E(S)=Y=Y_uY_q$. Finally, let ${\cal Y}_u, {\cal Y}_q$
be the share spaces for $U$ and $Q$ and let ${\cal Y}={\cal Y}_u \times {\cal Y}_q$ be the global share space.

Note that for the SS scheme to be perfect we must have
\begin{description}
\item[Correctness:] $H(S|Y_q)=0$. Equivalently, $S=f(Y_q)$ for some
  deterministic function $f$.
\item[Secrecy:] $I(S;Y_u)=0$. Equivalently,
  $P(Y_u=y_u|S=s)=P(Y_u=y_u|S=s')=P(Y_u=y_u) \quad \forall s,s' \in {\cal
    S}$.
\end{description}

Suppose now we have a quantum secret which is a linear superposition
of shares in ${\cal S}$ and a unitary map $\tilde{E}$ such that for $s
\in {\cal S}$:

$$
\tilde{E} \ket{s} = \sum_{y \in {\cal Y}} \sqrt{P(Y=y|S=s)} \ket{y}$$

This can in fact be rewritten as

\begin{multline*}
  \sum_{y_q : f(y_q)=s} \sqrt{P(Y_q=y_q|S=s)} \ket{y_q} \quad \cdot \\
  \sum_{y_u \in {\cal Y}_u} \sqrt{P(Y_u=y_u| Y_q=y_q)} \ket{y_u}
\end{multline*}

We want to decide if this is can correct erasures on $U$. To do so
requires showing that the density matrix of the $U$ component is
independent of the secret's state. Note that it is not sufficient to
show that the density matrix is the same for all $\ket{s}$. We have to
show this for all choices of the $\alpha_s$'s in $\sum_{s \in {\cal S}} \alpha_s
\ket{s}$. We can compute the density matrix explicitly by imagining
that a measure is made on the $Q$ component of the code and the
secret. We can then consider $P(S=s)$ to be $|\alpha_s|^2$. In what
follows \mbox{\boldmath$\rho_{U|y_q}$} is the density matrix of $U$
given $Y_q=y_q$.

\begin{multline*}
  \rho_u \\
  \shoveleft{ = \sum_{s \in {\cal S}} |\alpha_s|^2 \sum_{y_q \in {\cal Y}_q}
    P(Y_q=y_q|S=s) \mbox{\boldmath$\rho_{U|y_q}$}}\\
%   =  \sum_{s \in {\cal S},y_q \in {\cal Y}_q} P(S=s) P(Y_q=y_q|S=s)
%  {\textrm{density matrix of } U \choose \textrm{given } Y_q=y_q} \\
  \shoveleft{ =  \sum_{y_q \in {\cal Y}_q} P(Y_q=y_q) \mbox{\boldmath$\rho_{U|y_q}$}}\\
  \shoveleft{ =  \sum_{y_q \in {\cal Y}_q} P(Y_q=y_q) \cdot }\\
  \left( \sum_{y_u^{(1)} \in {\cal
        Y}_u} \sqrt{P(Y_u=y_u^{(1)} | Y_q=y_q)} \ket{y_u^{(1)}} \right) \cdot \\
  \left( \sum_{y_u^{(2)} \in {\cal Y}_u} \sqrt{P(Y_u=y_u^{(2)} |
      Y_q=y_q)} \bra{y_u^{(2)}}
  \right) \\
%   =  \sum_{y_u^{(1)},y_u^{(2)} \in {\cal Y}_u} \sum_{y_q \in {\cal Y}_q}
%  P(Y_q=y_q) \sum_{y_u^{(1)},y_u^{(2)} \in {\cal Y}_u} \\
%  \qquad \sqrt{P(Y_u=y_u^{(1)} | Y_q=y_q)P(Y_u=y_u^{(2)} | Y_q=y_q)}
%  \ket{y_u^{(1)}} \bra{y_u^{(2)}} \\
  \shoveleft{ = \sum_{y_u^{(1)},y_u^{(2)} \in {\cal Y}_u} \sum_{y_q \in
      {\cal Y}_q}
    \sqrt{P(Y_u=y_u^{(1)}, Y_q=y_q)} \cdot }\\
  \sqrt{P(Y_u=y_u^{(2)} ,Y_q=y_q)} \ket{y_u^{(1)}} \bra{y_u^{(2)}}
\end{multline*}

\noindent
The matrices in the set $$\left\{ \ket{y_u^{(1)}} \bra{y_u^{(2)}} :
  y_u^{(1)},y_u^{(2)} \in {\cal Y}_u \right\}$$
are linearly
independent. Their coefficients are

\begin{multline*}
  \sum_{y_q \in {\cal Y}_q} \sqrt{P(Y_u=y_u^{(1)},
    Y_q=y_q)P(Y_u=y_u^{(2)} ,Y_q=y_q)} \\
  \shoveleft{ = \sum_{s \in {\cal S}} |\alpha_s|^2 \sum_{y_q : f(y_q)=s}
    \sqrt{P(Y_u=y_u^{(1)},
      Y_q=y_q|S=s)}} \\
  \sqrt{P(Y_u=y_u^{(2)} ,Y_q=y_q|S=s)}
\end{multline*}

\noindent
For $\rho_u$ to be independent of the choice of $\alpha_s$ we must therefore
have
\begin{multline}
  \label{eq:ind}
  \sum_{y_q : f(y_q)=s} \sqrt{P(Y_u=y_u^{(1)}, Y_q=y_q|S=s)}\\
  \sqrt{P(Y_u=y_u^{(2)} ,Y_q=y_q|S=s)}
\end{multline}
independent of $s$ for all $y_u^{(1)},y_u^{(2)} \in {\cal Y}_u$. Thus

\begin{theorem}
  Given a classical SS scheme for an adversary structure ${\cal A}$,
  the correspnding quantum scheme corects erasures on $U \in {\cal A}
  \cap {\cal A}^*$, iff Equation (\ref{eq:ind}) is independent of $s$
  for all $y_u^{(1)},y_u^{(2)} \in {\cal Y}_u$.
\end{theorem}

As unnatural as this condition seems, it is nonetheless satisfied by
many SS schemes:

\begin{itemize}
\item If $Y_u$ is a function of $Y_q$ (as is the case in the Shamir
  scheme) then we have the expression (\ref{eq:ind}) equal to 0
  whenever $y_u^{(1)} \neq y_u^{(2)}$.  Furthermore, when $y_u^{(1)} =
  y_u^{(2)}=y_u$ the expression reduces to $\sum_{y_q : f(y_q)=s}
  P(Y_u=y_u,Y_q=y_q |S=s)$, which sums to $P(Y_u=y_u|S=s)$. This is
  independent of $s$ by the secrecy assumption above. Thus this type
  of scheme yields a secure QSS.
  
\item A group homomorphic secret sharing scheme is based on an injective
  homomorphism $h:G \times G^m\longrightarrow G^n$ for some group $G$. The
  secret $s$ is an element of $G$ and the $n$ shares are obtained by
  picking ${\bf v} \in_R G^m$ and calculating $h(s,{\bf v})$.
  
  In this case, the independence of expression (\ref{eq:ind}) from $s$
  is guaranteed by the following fact: in any homomorphic \textsc{ss}
  scheme, either two words $y_u^{(1)},y_u^{(2)}$ never appear with the
  same word $y_q$ (that is
  $$P(Y_u=y_u^{(1)}|Y_q=y_q) P(Y_u=y_u^{(2)}|Y_q=y_q)=0$$
  for all $y_q$) or they always appear with the same probability:
  \begin{multline*}
    \sqrt{P(Y_u=y_u^{(1)}|Y_q=y_q) P(Y_u=y_u^{(2)}|Y_q=y_q)} \\
    = P(Y_u=y_u^{(1)}|Y_q=y_q).
  \end{multline*}
  
  The same analysis as before applies: \textsc{qss} schemes constructed from
  homomorphic schemes are secure. Interestingly, there seem to be no
  cases where non-homomorphic schemes provide any advantage over
  homomorphic ones \cite{StinsonPersonal}.
\end{itemize}

Thus, it seems that although not all classical \textsc{ss} schemes
yield a \textsc{qss} scheme directly, the most important ones do.
However, the proof given does not give the reconstruction procedure;
it only proves its existence. It is not \emph{a priori} clear that all
classical SS schemes which yield a secure QSS scheme will have
efficient (quantum) reconstruction procedures.

\subsection*{Acknowledgements}
\label{sec:ack}

I would like to thank Richard Cleve, Claude Cr{\'e}peau, Daniel Gottesman
and Paul Dumais for helpful discussions.

\end{document}